\documentclass{article}
\usepackage[utf8]{inputenc}
\usepackage[english]{babel}

\usepackage{graphicx}
\usepackage[space]{grffile}
\usepackage{latexsym}
\usepackage{authblk}
\usepackage{textcomp}
\usepackage{longtable}
\usepackage{tabulary}
\usepackage{booktabs,array,multirow}
\usepackage{amsfonts}
\usepackage{amsmath}
\usepackage{amssymb}
\usepackage{amsthm}
\usepackage{natbib}
\usepackage{url}
\usepackage{soul} % added by the author
\usepackage{todonotes} % added by the author
\usepackage{lineno} % added by the author
\usepackage{comment} % added by the author
\usepackage{hyperref}
\usepackage[title]{appendix} % For appendix name

\newcommand{\hidefigure}{1} % 0 = hide figures, 1 = add figures
\renewcommand{\hl}[1]{#1} % hide highlights
\newtheorem{definition}{Definition}
\newtheorem{theorem}{Theorem}

\title{Using the rejection sampling for finding tests}

\author[1]{Markku Kuismin}
\date{}

\affil[1]{Research Unit of Mathematical Sciences, University of Oulu, Oulu, Finland, markku.kuismin@oulu.fi}

\begin{document}

\maketitle
\selectlanguage{english}
\begin{abstract}
A new method based on the rejection sampling for finding statistical tests is proposed. This method is conceptually intuitive, easy to implement, and applicable for arbitrary dimension. To illustrate its potential applicability, three distinct empirical examples are presented: (1) examine the differences between group means of correlated (repeated) or independent samples, (2) examine if a mean vector equals to a specific fixed vector, and (3) investigate if samples come from a specific population distribution. The simulation examples indicate that the new test has similar statistical power as uniformly the most powerful (unbiased) tests. Moreover, these examples demonstrate that the new test is a powerful goodness-of-fit test.
\end{abstract}%

\section{Introduction}\label{1sec}

Statistical hypothesis testing is a cornerstone of statistical inference. Numerous tests have been developed for various hypothesis testing problems, such as assessing statements about group means, testing for statistical independence \citep[see, e.g.,][]{Reshef.etal2018, Kuismin2025}, assess whether two samples came from the same underlying distribution \citep[see, e.g.,][]{Fasano&Franceschini1987, Chen.etal2025}, test the normality of the data \citep[see, e.g.,][]{Zhou&Shao2014} or determining whether two categorical variables are associated, to mention just a few examples from the extensive body of statistical tests \citep[see, e.g.,][]{Lehmann&Romano2005, Kanji2006}.

The key goal in developing statistical tests is to create methods that can always distinguish genuine effects from those that arise purely by chance. In practice, however, a more realistic objective is to have a toolbox of tests that minimize the risk of erroneous conclusions about the population while maintaining a high probability of detecting true effects, provided that the specific characteristics of the problem at hand are carefully taken into account. The development of such statistical tests remains a fundamental problem in statistical research and continues to be an active area of investigation.

An addition to this toolbox is introduced in this paper. In particular, a new method for constructing statistical tests is proposed which is based on the rejection sampling commonly used to generate random samples from a probability distribution of interest. In sum, the current study aims to demonstrate how to use the probability of acceptance of the rejection sampling as a test statistic and provide a conceptual overview of the proposed approach. It is demonstrated that this method is versatile, easy to use, and exhibits high statistical power. The new method is used to derive tests for the following problems: (1) compare means of two independent or correlated variables; (2) examine the mean vector of a multivariate data of independent or correlated variables; and (3) investigate if samples come from a specific univariate or multivariate population distribution (goodness-of-fit). The performance of the tests derived using the new method is evaluated through Monte Carlo (MC) simulations to assess their statistical power. These examples suggest that this new framework demonstrates high statistical power as state-of-the-art tests, and in case (3) described above it exhibits even higher statistical power compared to the state-of-the-art tests. 

\section{Materials and methods}\label{sec:materials&methods}
\subsection{Proposed framework}

Following \cite{Casella&Berger2024}, \hl{let $\theta$ denote a population quantity of interest (which may be a parameter vector or a more general functional) taking values in a space} $\Theta$. Let $\Theta_0$ be some subset of $\Theta$, and let $\Theta_0^c$ denote its complement. Throughout this paper, $H_0$ and $H_A$ are used to denote the null and alternative hypotheses, respectively. In the parametric setting, the general format of null and alternative hypothesis is $H_0:\ \theta \in \Theta_0 \quad \text{against} \quad H_A:\ \theta \in \Theta_0^c$. For example, one might be interested in testing $H_0:\ \theta = 0 \quad \text{against} \quad H_A:\ \theta \neq 0$ where $\theta$ is a population parameter.

There are two types of errors a researcher can do when making a decision whether or not reject $H_0$: Type I error (reject $H_0$ when it is actually true) and Type II error (accept $H_0$ when it is not true). The probabilities corresponding to these errors can be expressed informally as follows: Use $\alpha = P(\text{Reject } H_0 \mid H_0 \text{ is actually true})$ to denote the Type I error, and let $\beta = P(\text{Accept } H_0 \mid H_0 \text{ is actually false})$ denote the Type II error. A typical practice is to evaluate a test in terms of the statistical power instead of the Type II error. The statistical power of a test is defined as $1 - \beta$. An ideal test would have both $\alpha$ and $\beta$ as zero. Except in simple situations, this ideal cannot be attained and these errors can never be truly zero because there is always a chance, even if small, of observing data that falsely leads to the rejection of a true (false) null hypothesis due to random variation. In practice, a good test has statistical power close to one for most $\theta \in \Theta_c$, while maintaining the Type I error at a pre-specified level $\alpha \in [0, 1]$ that is close to zero for most $\theta \in \Theta_0$.

Let $X = (x_1,\ldots,x_p)^\top$ be a $p$-dimensional real-valued random vector, and let $\mathbf{X} = (X_1,\ldots,X_n)$ denote the corresponding data matrix consisting of $n$ independent observations. A statistical hypothesis test is defined using a test statistic $T(X_1, \ldots, X_n) = T(\textbf{X})$. In most cases, the test says that $H_0$ should be rejected if $T(\textbf{X})$ is greater than (less than) or equal a certain constant $c$. This constant $c$ is called the rejection threshold, and the rejection region can be written as: $\{\textbf{X} \mid T(\textbf{X}) \geq c\}$ where $c \in \mathbb{R}$.

There are different ways for finding test to assess if the observed data provides any evidence against $H_0$. Three of the most popular, well-known and studied methods are Wald test, the score test (also known as Lagrange multiplier), and the Likelihood Ratio (LR) test. From these methods the new framework presented in this paper is related to the LR test defined as follows:

\begin{definition}\label{def:LR_stat}
Let $\textbf{X} = (X_{1}, \ldots, X_n)$ be a random sample (data matrix) from a population with probability density function or probability mass function $f(\textbf{X} \mid \theta)$. The likelihood function is the joint density (or mass) function $L(\theta \mid \textbf{X}) = f(\textbf{X} \mid \theta)$. The Likelihood Ratio test statistic for testing $H_0:\theta \in \Theta_0$ against $H_A:\theta \in \Theta_0^c$ is 

$$\lambda(\textbf{X}) = \frac{\sup_{\Theta_0} L(\theta \mid \textbf{X})}{\sup_{\Theta}L(\theta \mid \textbf{X})}.$$
A LR test is any test that has a rejection region of the form $\{\textbf{X} \mid \lambda(\textbf{X}) \leq c\}$, where $0 \leq c \leq 1$.

\end{definition}

In the next section it is shown how the rejection sampling, also known as the accept-reject algorithm can be used to find statistical test. First, the basic principles of the AR algorithm are briefly reviewed to provide the necessary background, allowing the reader to follow the ideas presented in this paper.

\subsection{The accept-reject algorithm}

Rejection sampling, also known as the accept-reject algorithm (hereafter AR algorithm), is one of the conceptually simplest algorithms used to generate samples from arbitrary probability distributions of any dimension \citep[see, e.g.,][pp. 51--57]{Robert&Casella2010}. Let $f$ be a function presenting the functional form of the density of interest. For simplicity, $f$ is referred as the probability density function (pdf) of the distribution of interest; the algorithm is identical in case of probability mass function (pmf). $f$ is sometimes called as the target distribution. Use $X \sim f$ to indicate that the variable $X$ follows a specific but unknown multivariate probability distribution with density $f$. Let $g$ denote a known univariate or multivariate probability distribution with known pdf ($g$ is sometimes called as the proposal distribution). Let $D > 0$ be a constant such that $f(X) \leq Dg(X)$ The AR algorithm can be outlined into the following four steps:

\begin{enumerate}
    \item Generate $X_i = (x_{1i}, \ldots, x_{pi})^\top \sim g$ and $u_i \sim Unif(0, 1)$.
    \item Compute $a = f(X_i)/[Dg(X_i)]$.
    \item If $a > u_i$, accept $X_i$ as a sample from $f$.
    \item Repeat steps 1.- 3. until the desired number of pseudo-random samples has been obtained.
\end{enumerate}

The empirical efficiency of the AR algorithm can be evaluated using the so-called probability of acceptance, which is defined as follows: Let $\text{I}(x)$ denote the indicator function\hl{/random variable} of the event $x > u$, where $u \sim \text{Unif}(0, 1)$. It is assumed that $N \geq 1$ rounds of rejection sampling are performed. The probability of acceptance is given as the ratio of accepted samples to the total number of simulation rounds $N$, more formally:

\begin{equation}\label{eq:prob_accept}
    \rho = \frac{1}{N}\sum_{i=1}^N \text{I}\Bigl(\frac{f(X_i)}{Dg(X_i)}\Bigr).
\end{equation}
The probability of acceptance, $0 \leq \rho \leq 1$, can be used to evaluate how closely the proposal distribution $g$ matches the target distribution. It will be shown, that with small modifications the rejection sampling provides a very intuitive framework for deriving statistical tests with a straightforward interpretation (see theorem \ref{theorm:rho_eq_e}).

\subsection{Deriving a test using the principles of rejection sampling}

To elaborate how the underlying principle of the AR algorithm is applied to find tests, consider a practical goodness-of-fit problem \hl{for multivariate distributions}: $H_0:f(X_i) = f_0(X_i)$ for all $X_i$, $i=1,\ldots,n$ where $f$ is a population pdf and $f_0$ is a hypothesized pdf, e.g., multivariate normal density. Compare this to (\ref{eq:prob_accept}) and draw a parallel between $f_0$ and the target distribution of the AR algorithm. Under the null hypothesis, the density $g$ from which the data are drawn should agree with the density $f_0$. Instead of generating pseudo-random numbers from $g$, the observed samples $X_1, \ldots, X_n$ are used as input to the algorithm. The likelihood ratios $f_0(X_i)/g(X_i)$, $i = 1, \ldots, n$, can then be used to assess the degree of agreement between the hypothesized pdf $f_0$ and density $g$. \hl{Let $g$ be a density estimate and assume, that $H_0$ is true. In this special case}, \textit{the resulting probability of acceptance (\ref{eq:prob_accept}) should equal one if the estimated density matches with a given probability distribution $f_0$}; setting $D = 1$ represents an uncommon case when using the AR algorithm but a logical choice when investigating if the data provides sufficient evidence against the specified theoretical distribution $f_0$.

This kind of test provides a conceptually intuitive interpretation: the probability of acceptance literally measures how often one would ``accept'' the observed sample under $H_0$; equivalently, it can be formulated as a bounded likelihood-ratio statistic where one of the likelihoods is empirical induced by a randomization step. In the next subsections, two example tests derived from the AR framework are considered, and some of their theoretical properties are elaborated.

\subsubsection{AR goodness-of-fit-test}

The first example test is a goodness-of-fit test, which was already discussed informally in the previous section. More formally, the basic form of the test statistic can be defined as follows:

\begin{definition} \label{def:ar_stat_gof}
    Let $f_0$ be the claimed population pdf $f$ under $H_0: f = f_0$. Let $\text{I}(x_i)$ be the indicator function of the event $x_i > u_i$ where $u_i \sim Unif(0, 1)$. Let $T(\mathbf{X}) = n^{-1} \sum_{i=1}^n \text{I}[f_0(X_i)/\widehat{f}(X_i)]$ be the mean of indicator variables depending on the ratios between the theoretical density $f_0$ and density estimate $\widehat{f}$, evaluated at the sample values. Use the statistic $T(\textbf{X})$ with rejection region $\{\textbf{X} \mid T(\textbf{X}) \leq c\}$, where  $0 \leq c \leq 1$ for testing $H_0:f = f_0$ against $H_A:f \neq f_0$.
\end{definition}

However, there is one problem: $T(\mathbf{X})$ depends on an external random variable $U$, meaning that its value may change even if the ratios $f(X_i)/g(X_i)$, $i = 1,\ldots,n$, remain unchanged. Therefore, instead of using $T(\mathbf{X})$ directly as a test statistic, the expected value of $T(\mathbf{X})$ is used to evaluate the evidence that the data provide against (or for) $H_0$. The AR test statistic is defined as follows:

\begin{definition} \label{def:ar_stat_gof_e}
Let $E(Y)$ be the expected value of a random variable $Y$. Let $T(\mathbf{X}) = n^{-1} \sum_{i=1}^n \text{I}[f(X_i)/\widehat{f}(X_i)]$. The AR test statistic is the expected value of $T(\textbf{X})$ with respect to $U \sim Unif(0, 1)$, $\rho(\textbf{X}) = E_U[T(\mathbf{X})]$. Use the statistic $\rho(\textbf{X})$ with rejection region $\{\textbf{X} \mid \rho(\textbf{X}) \leq c\}$, where  $0 \leq c \leq 1$ for testing $H_0:f = f_0$ against $H_A:f \neq f_0$.
\end{definition}

One could compute $\rho(\mathbf{X})$ by averaging results over multiple runs of the AR test, but this increases the computational complexity of the method. However, the next theorem shows that such time-consuming simulations can be avoided and that computing the expectation $\rho(\mathbf{X})$ is straightforward.

\begin{theorem}\label{theorm:rho_eq_e}
    Let $r_i = f_0(X_i)/\widehat{f}(X_i)$. Let $T(\boldsymbol{X})$ be defined as in Definition \ref{def:ar_stat_gof} and \ref{def:ar_stat_gof_e}. The expected values of $T(\boldsymbol{X})$ is:
    
\begin{equation}\label{eq:ar_stat}
    \rho(\textbf{X}) = E_U\left[T(\textbf{X})\right] = n^{-1}\sum_{1=1}^n \min\left(1, f_0(X_i)/\widehat{f}(X_i)\right).
\end{equation}

\end{theorem}
\noindent \textit{Proof.} See the Appendix.

The following theorem shows that the AR goodness-of-fit test is consistent against any fixed alternative. In particular, the AR test is directly related to the total variation distance between the null and the true distribution. This contrasts with likelihood ratio and empirical likelihood tests, whose asymptotic behavior is governed by the Kullback–Leibler divergence \citep[see, e.g.,][]{Eguchi&Copas2006}.

\begin{theorem}\label{theorm:rho_eq_tvd}
Let $X_1,\dots,X_n \in \mathbb{R}^p$ be i.i.d. random vectors with common density $f$ and $\textbf{X} = (X_1, \ldots, X_n)$. Let $f_0$ be a fixed null density on $\mathbb{R}^p$. Suppose that $f$ and $f_0$ are strictly positive and bounded away from zero and infinity on their common support $\mathcal{X}$ and that the density estimator $\widehat{f}$ satisfies

\begin{align}
        \sup_{X \in \mathcal{X}} \big| \widehat{f}(x) - f(x) \big|
        \xrightarrow{P} 0.
\end{align}
The total variation distance (TVD) between $f(x)$ and $f_0(x)$ is defined as 
\begin{align}
\|f(x) - f_0(x)\|_{\text{TV}} = \frac{1}{2}\int_{\mathcal{X}}|f(x) - f_0(x)|\, dx.
\end{align}
Then, as $n \to \infty$,

\begin{align}
    \rho(\textbf{X}) \xrightarrow{P} 1 - \|f(x) - f_0(x)\|_{\text{TV}}.
\end{align}
\end{theorem}
\noindent \textit{Proof.} See the Appendix.

From Theorem~\ref{theorm:rho_eq_tvd} it is easy to see that the value of the test statistic is always between zero and one. In particular, when $H_0$ is true, the value of the test statistic approaches one as the sample size increases. On the other hand, when $H_0$ is not true and the sample size increases, the value of the test statistic approaches a value strictly smaller than one, determined by the TVD between the population density $f$ and the theoretical density $f_0$. Consequently, the power of the AR test depends on the sample size and how accurately the population density can be estimated. The relationship between the AR goodness-of-fit test and the TVD is illustrated in Figure \ref{fig:tvd}.

\begin{figure}[ht]
\begin{center}
    \if1\hidefigure
{
    \includegraphics[width=0.9\columnwidth]{tvd_both.png}
}\fi
\caption{{The asymptotic behavior of the AR goodness-of-fit test statistic when $f_0$ is the probability density function of the univariate standard normal distribution and $n \in \{100, 500, 1000, 1500, 2000\}$. The lines represent the values of $\rho(\mathbf{X})$ averaged over 100 simulations. In the left panel, the true population distribution is a univariate $t$-distribution with $df = 6$ ($H_0$ is false). In the right panel, the true population distribution is the univariate standard normal distribution ($H_0$ is true). The horizontal dashed line illustrates the true value $1 - \|f(X) - f_0(X)\|_\text{TV}$. The blue line corresponds to the value of the test statistic $\rho(\mathbf{X})$ computed using the true population density $f$. The red line corresponds to the values of the test statistic when $f$ is estimated using a kernel density estimator (KDE). Vertical bars represent the 95\% confidence interval estimates.
{\label{fig:tvd}}%
}}
\end{center}
\end{figure}

\subsubsection{AR test for a parameter vector}

The second application of the AR test considered in this paper is to test values of a population parameter vector $\theta = (\theta_1, \ldots, \theta_q)^\top$, $q \ge 1$. Since the construction parallels the goodness-of-fit AR test, the resulting procedure inherits its theoretical properties and is defined as follows:

\begin{definition} \label{def:ar_stat_p}
    Let $f_0$ be the pdf under $H_0: \theta \in \Theta_0$, $g$ be the pdf of $\widehat{\theta}$, and $T(\mathbf{X}) = n^{-1} \sum_{i=1}^n \text{I}[f_0(X_i\mid \theta_0)/g(X_i\mid \widehat{\theta})]$ be the mean of indicator variables depending on the ratios between $f_0$ with respect to the parameter values under $H_0$ and a density $g$ with respect to the estimated parameters values. Use the AR test statistic (\ref{eq:ar_stat}) with rejection region $\{\textbf{X} \mid \rho(\textbf{X}) \leq c\}$, where  $0 \leq c \leq 1$ for testing $H_0:\theta \in \Theta_0$ against $H_A:\theta \in \Theta_0^c$.
\end{definition}

The AR test is not limited to tests described in Definitions \ref{def:ar_stat_gof_e} and \ref{def:ar_stat_p}; with small modifications it can be used in many testing problems. For example, in \citet{Kuismin2025} the author used a similar test to examine if two univariate random variables are statistically independent. The definition of the AR test is flexible: in principle, it can be defined with respect to different assumptions about the proposal distribution. For instance, while testing claims about the population covariance matrix \citep[see, e.g.,][]{Bodnar.etal2025}, one could use the multivariate normal distribution, inverse Wishart distribution, the Lewandowski-Kurowicka-Joe distribution \citep{Lewandowski.etal2009}, or define it as a ratio of two density estimates in a two-sample testing problem.

\subsection{The distribution of the statistic $T(\textbf{X})$}

Estimating the distribution of $T(\textbf{X})$ by simulation is straightforward: By running the AR algorithm multiple times, the resulting values of $T(\textbf{X})$ can be used to approximate its distribution with respect to the data $\textbf{X}$. This allows evaluating the evidence provided by the AR test without resampling the data. The resulting interval estimates are credible intervals with a clear probabilistic interpretation. Moreover, it can be shown that the statistic $nT(\textbf{X})$ follows a Poisson binomial distribution. For simplicity, consider the case when the goodness-of-fit test, as defined in Definition \ref{def:ar_stat_gof}, is applied.

Let $b_i = \text{I}[f_0(X_i)/\widehat{f}(X_i)]$, $i = 1, \ldots, n$. From the properties of the Bernoulli distribution and the definition of the Poisson binomial distribution, it follows that

\begin{equation}\label{eq:poi_bin}
    nT(\textbf{X}) = \sum_{i=1}^n b_i \sim \text{PoiBin}(\rho_1, \ldots,\rho_n),
\end{equation}
where $\text{PoiBin}(\rho_1, \ldots, \rho_n)$ denotes a Poisson binomial distribution with parameters $\rho_i = P(b_i > U) = \text{min}\{1, f_0(X_i)/\widehat{f}(X_i)\}$, $i = 1, \ldots, n$. When $\rho_1 = \cdots = \rho_n = \rho'$, the Poisson binomial distribution reduces to the binomial distribution $\text{Bin}(n, \rho')$.

%From the properties of the indicator function and the central limit theorem, it follows that the AR test statistic converges in distribution to $N(\mu_{\rho}, \sigma_{\rho}^2)$ distribution, where $\mu_{\rho} = \rho(\textbf{X})$ and $\sigma_{\rho}^2 = \rho(\textbf{X})(1 - \rho(\textbf{X}))/n$.

\begin{figure}[ht]
\begin{center}
    \if1\hidefigure
{
    \includegraphics[width=0.9\columnwidth]{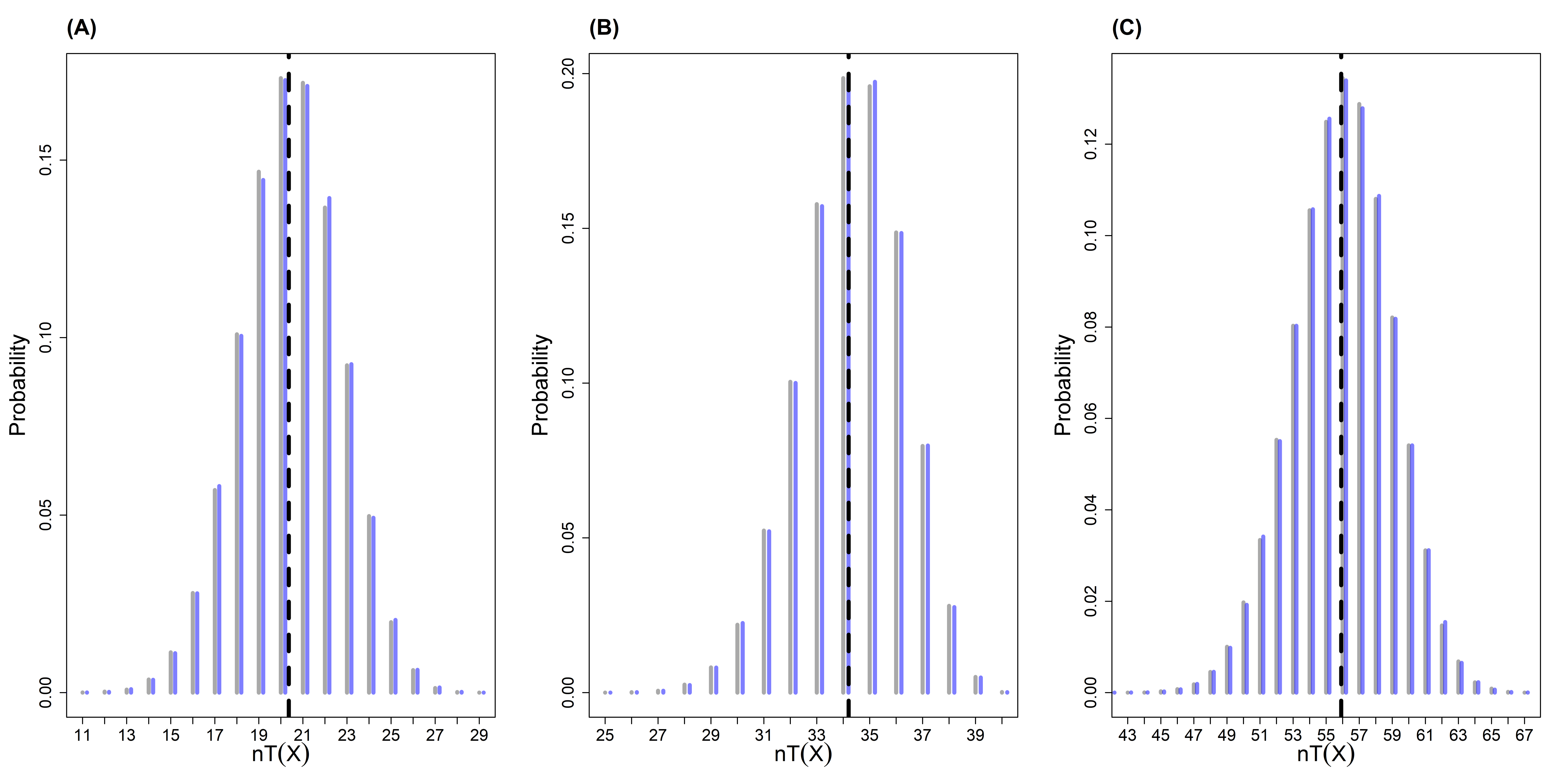}
}\fi
\caption{{Probability mass function estimate of the statistic $nT(\textbf{X})$ based on 10000 simulation replicates (gray bars) and the Poisson binomial distribution (blue bars) when the AR test is used to test univariate normality (see Definition \ref{def:ar_stat_gof} and Section \ref{sec:goodness-of-fit}). Different panels represent different sample sizes: \textbf{(A)} 30; \textbf{(B)} 40; and \textbf{(C)} 70. The vertical red dashed line illustrates the value $n\rho(\textbf{X})$ and the vertical solid line illustrates the value $n(1 - ||f(x) - f_0(x)||_{TV})$.
{\label{fig:sampling_dist}}%
}}
\end{center}
\end{figure}

In Figure \ref{fig:sampling_dist}, the simulated and exact distributions of $nT(\textbf{X})$ are compared when the AR test is applied to test univariate normality (a detailed description of this example is provided in Subsection \ref{sec:goodness-of-fit}). In Figure \ref{fig:sampling_dist}, the Poisson binomial distribution and the simulated distribution of $nT(\textbf{X})$ are shown to be virtually identical. Consequently, the percentiles of the Poisson binomial distribution can be used to compute credible intervals for the statistic $T(\textbf{X})$ with respect to the data.

The null distribution of test statistic $\rho(\textbf{X})$ must be specified in order to determine the rejection threshold $c$ and the $p$-value of the AR test. A natural choice is $c=1$, but this makes the test too liberal since $H_0$ would be rejected for any $\rho(\textbf{X}) < 1$. In addition, the null distribution is not well-defined in this case because the variance of the test statistic is zero. Thus, setting $c=1$ is not appropriate for any application. 

From the central limit theorem, it follows that the null distribution of $\rho(\textbf{X})$ is approximately normal. However, this normal approximation typically provides a crude estimate of the null distribution, since $\rho(\textbf{X})$ is close to one under $H_0$. Moreover, when applying, for instance, the goodness-of-fit AR test, one should take into account the potentially slow convergence of the density estimator $\widehat{f}$ to the true population density $f$ (see Figure \ref{fig:tvd}). For this reason, the null distribution is estimated using a simulation approach. In particular, a Monte Carlo procedure is used to determine the rejection threshold $c$ and, consequently, the $p$-value of the AR test \citep[see, e.g.,][for Monte Carlo significance testing]{Howes2023}. The following Monte Carlo significance testing procedure is used in this paper \hl{(where a small value of $\rho(\textbf{X})$ heuristically indicates evidence against $H_0$)}:

\begin{enumerate}
    \item Generate $n$ independent and identically distributed (i.i.d.) values $\tilde{\textbf{X}} = (\tilde{X}_1, \ldots, \tilde{X}_n)$ from the target distribution $f_0$ under $H_0$.
    \item Calculate value of the test statistic $\rho(\tilde{\textbf{X}})$.
    \item Repeat steps 1 to 2 $M$ times to get $\rho(\tilde{\textbf{X}})_1, \ldots, \rho(\tilde{\textbf{X}})_M$ of the AR test statistic.
    \item Use $\rho(\tilde{\textbf{X}})_1, \ldots, \rho(\tilde{\textbf{X}})_M$ to estimate the null distribution of $\rho(\textbf{X})$ and calculate the Monte Carlo $p$-value, $p_{ar} = [\sum_{m=1}^M\text{I}(\rho(\tilde{\textbf{X}})_m \leq \rho(\textbf{X})) + 1]/(M + 1)$.
\end{enumerate}

The percentile (rejection threshold) of the null distribution of $\rho(\mathbf{X})$ can be approximated by estimating the parameters of the Poisson binomial distribution with only a small number of Monte Carlo simulations. Table \ref{tab:qest} shows percentiles of $\rho(\mathbf{X})$ based on 10000 Monte Carlo simulations and on the Poisson binomial distribution $\text{PoiBin}(\widehat{\rho}_1,\ldots,\widehat{\rho}_n)$, where $\widehat{\rho}_i$, $i = 1,\ldots,n$, as defined in (\ref{eq:poi_bin}), are estimated under $H_0$ as the sample means from just 50 Monte Carlo replicates. The differences between percentiles obtained from the two approximations are negligible, and either method can be used without practical impact on inference as the sample size increases.

\begin{table}[!ht]
\begin{center}
\begin{tabular}{l|l|cc|cc}
\hline
 & & \multicolumn{2}{c|}{AR goodness-of-fit test} & \multicolumn{2}{c}{AR group mean difference test} \\ \hline
$n$ & Percentile & PoiBin approx. & MC & PoiBin approx. & MC \\ \hline
10 & 0.01 & 0.600 & 0.586  &  0.600 & 0.517 \\ 
   & 0.05 & 0.700 & 0.719  &  0.700 & 0.660 \\
   & 0.10 & 0.800 & 0.777  &  0.700 & 0.719 \\ \hline
30 & 0.01 & 0.800 & 0.803  &  0.767 & 0.763 \\ 
   & 0.05 & 0.867 & 0.852  &  0.800 & 0.820 \\
   & 0.10 & 0.867 & 0.878  &  0.833 & 0.849 \\ \hline
60 & 0.01 & 0.883 & 0.863  &  0.850 & 0.841 \\ 
   & 0.05 & 0.900 & 0.899  &  0.867 & 0.876 \\
   & 0.10 & 0.917 & 0.915  &  0.883 & 0.893  \\
\end{tabular}
\end{center}
\caption{{Percentiles of the empirical null distribution of $\rho(\textbf{X})$, estimated using the Poisson binomial distribution approximation and Monte Carlo (MC) simulations, for the AR-based goodness-of-fit test (Definition \ref{def:ar_stat_gof_e}) and group mean difference test (Definition \ref{def:ar_stat_p}).} \label{tab:qest}}
\end{table}

In the following sections, the empirical statistical power of three distinct AR tests is assessed through a comparative Monte Carlo simulation study, where the rejection threshold $c$ is determined by Monte Carlo simulations. \hl{It will be demonstrated numerically that, when $H_0$ is true, the probability that the AR tests incorrectly reject $H_0$ does not exceed the nominal significance level. Moreover, when a real effect is present, the power of the AR tests approaches one as the sample size increases. Each new AR test is also compared with state-of-the-art statistical tests.}

\section{Simulations}

\subsection{Comparing equality of group means} \label{sec:gmean}

This first example \hl{illustrates the parametric use of the AR test framework}. In particular, the difference between group means is investigated, which represents one of the fundamental problems in statistical inference dating back to the pioneering work of \citet{Student1908}. Let $\mu = (\mu_1, \ldots, \mu_p)^\top$ denote a population mean vector of length $p \geq 2$. Consider testing:

\begin{align}\label{hypo:paired_hypothesis}
    H_0&: \mu_1 =\mu_2=\ldots=\mu_p \nonumber \\
    H_A&: \mu_i \neq \mu_j \quad \text{ for some } i \text{ and } j \text{ where } i \neq j. \nonumber
\end{align}

Here the AR test defined in Definition \ref{def:ar_stat_p} is applied for testing the hypothesis described above. In this example, it is demonstrated how incorporating certain distributional assumptions can improve the computational efficiency of the AR test. It is illustrated how the AR test can be constructed using a sufficient statistic. Here, the test is constructed based on the sample mean vector, which is a sufficient statistic for $\boldsymbol{\mu}$. The AR test for the sample mean vector is defined as follows: Let $\bar{X} = [\sum_{i=1}^n x_j^i/n]_{p \times 1}$ be the sample mean vector of length $p$. Denote $\mu_0 = (\mu_1^0, \ldots, \mu_p^0)^\top$ be a parameter vector of length $p$ under $H_0$. Examine the value of the following statistic

\begin{equation} \label{eq:mvclt}
\textbf{t} = \sqrt{n}(\bar{X} - \mu_0).
\end{equation}
According to the multivariate central limit theorem, $\textbf{t}$ converge in distribution to multivariate normal distribution $N(\textbf{0}, \Sigma)$ under the null hypothesis where \textbf{0} is a zero vector of length $p$. Thus, the pdf of $N(\textbf{0}, \Sigma)$ is used as the target distribution. Here a multivariate $t-$distribution is used as a proposal distribution to evaluate if the sample mean vector values differ from each other. Let $f_n \sim N(\textbf{0}, \Sigma)$ and $g_n \sim t(n-1, \textbf{0}, \Sigma)$ where $f_n$ and $g_n$ denote the pdf of the multivariate normal and multivariate $t$-distributions, respectively. Let $T(\textbf{t}) = \text{I}[f_n(\textbf{t})/g_n(\textbf{t})]$ denote the indicator variable depending on the ratio between $f_n$ and $g_n$ under $H_0$. Clearly $f_n(\textbf{t})/g_n(\textbf{t}) > 0$ for all $\textbf{t} \in \mathbb{R}^p$ and $g_n(\textbf{t}) \neq 0$. Under the null hypothesis set $\mu_0 = [a]_{p\times 1}$ where $a = \sum_{j = 1}^p \sum_{i=1}^n x_i^j/(pn)$. In this special case the AR test statistic simplifies to

\begin{equation}\label{eq:rhoteststat}
    \rho(\textbf{X}) = E_U\left[T(\textbf{t}) > U\right] = P\left[T(\textbf{t}) > U\right] = \min\{T(\textbf{t}), 1\}.
\end{equation}
Now $f_n$ and $g_n$ depend on an unknown population parameter $\Sigma$. The sample covariance matrix $\widehat{\Sigma} = \sum_{i=1}^n (X_i - \bar{X})(X_i - \bar{X})^\top/(n-1)$ is employ in place of $\Sigma$.

The empirical power of the AR test is compared with the paired $t$-test, two-sample $t$-test, and the Likelihood Ratio (LR) test when $p = 2$. According to the Neyman-Pearson Lemma, the LR test is uniformly the most powerful (UMP) test when considering a simple hypothesis $H_0: \mu_d = 0$ against $H_A: \mu_d = \eta$ where $\mu_d = \mu_1 - \mu_2$ and $\eta$ is the true difference between group means. The LR test is based on the assumption that all parameter values are known and does not involve parameter estimation. It is included here as a baseline test to illustrate the power that the optimal test could have when $p=2$ \citep[see, e.g.][pp.~387--394]{Casella&Berger2024}. The sample size $n$ is set as 52 which is the minimal number of observations needed for the two sided paired $t$-test to have power at least $0.8$ when assuming that the group variances are equal and the effect size (Cohen's D) is small \citep{Cohen1988}. 

\begin{figure}[ht]
\begin{center}
    \if1\hidefigure
{
    \includegraphics[width=0.9\columnwidth]{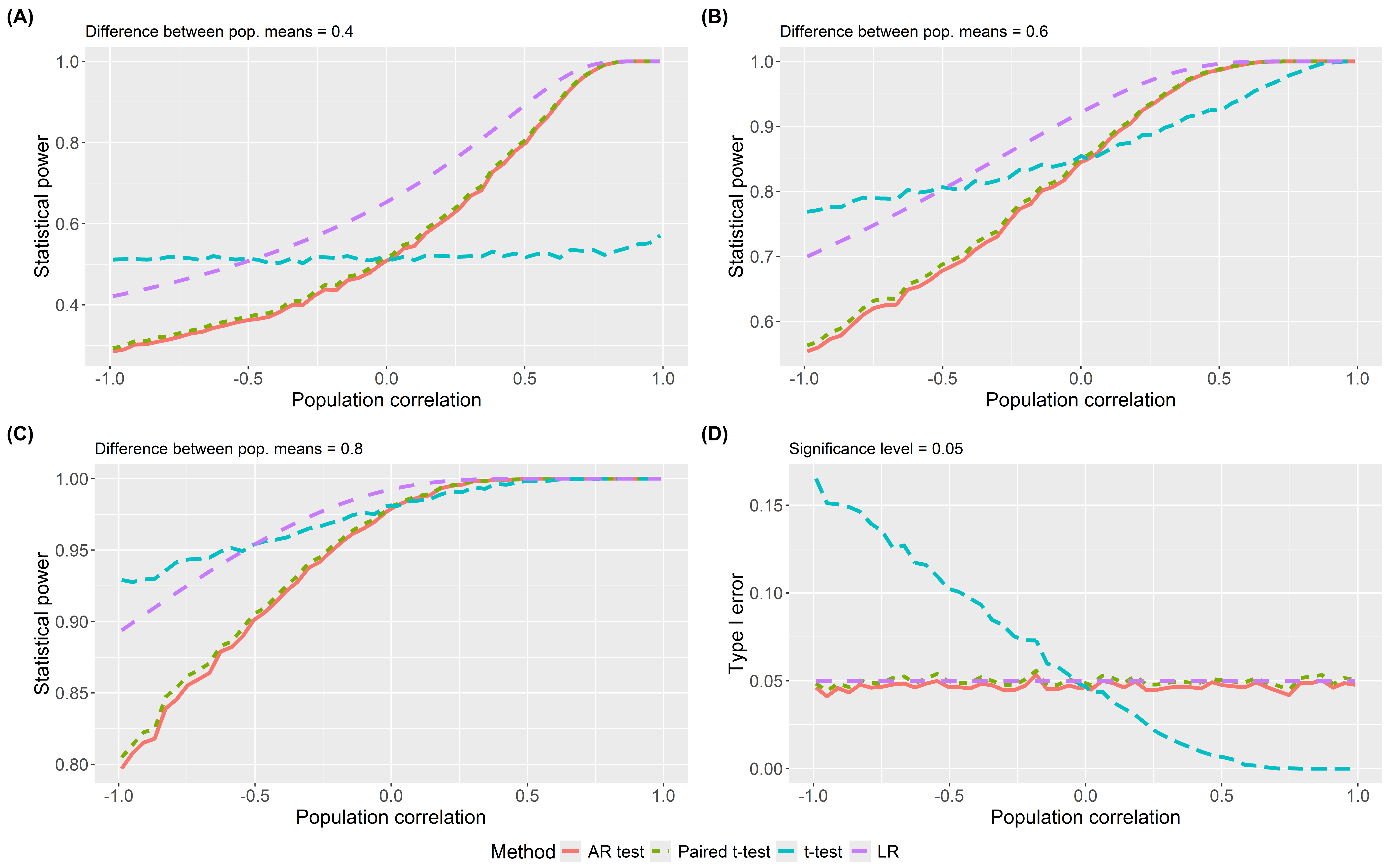}
}\fi
\caption{{Population correlation vs. the empirical power of the AR test (red solid line),  Likelihood Ratio test (LR) (dotted purple line), paired $t$-test (solid red line), and independent sample $t$-test ($t$-test) (dashed blue line) at $\alpha = 0.05$. Different panels illustrate the power vs. correlation when the true difference between two group means is: \textbf{(A)} 0.4; \textbf{(B)} 0.6; and \textbf{(C)} 0.8. Panel \textbf{(D)} illustrates the estimated Type I error associated with different tests. The predetermined significance level 0.05 is illustrated with horizontal dashed line in panel \textbf{(D)}. Here $n = 52$.
{\label{fig:norm_power}}%
}}
\end{center}
\end{figure}

For this simulation example, Let $X = (x_1, x_2)^\top \sim N(\mu, \Sigma)$ where $\mu = (\mu_1, \mu_2)^\top$ and $\Sigma = [\sigma_{ij}]$, $i,j \in \{1, 2\}$ is a symmetric and positive definite matrix. Consider the hypothesis test $H_0: \mu_1 = \mu_2$ against $H_A: \mu_1 \neq \mu_2$. Here it is assumed that group variances are equal, $\sigma_{ii} = 1$ and $\sigma_{ij} \in [-0.99, 0.99]$. The power for each test is estimated when the difference between population mean parameters change. Let $\eta = \mu_1 - \mu_2$ when $\eta \in \{0.4, 0.6, 0.8\}$ at different values of the population correlation $\sigma_{ij}$. The empirical power of different tests is the ratio of significant tests of 10000 Monte Carlo samples at $\alpha = 0.05$. Figure \ref{fig:norm_power} illustrates the empirical power with respect to the population correlation.

The power of the AR test is slightly lower compared to the paired $t$-test. However, from Figure \ref{fig:norm_power} (\textbf{D}) one can see that the Type I error of the AR test as defined in (\ref{eq:rhoteststat}) is below the predetermined level of significance. It appears that using a heavy-tailed distribution as the proposal distribution results in a conservative test. Figure \ref{fig:norm_power} shows that when there is a strong positive, linear association between variables (groups) the statistical power of all methods increases.

\begin{figure}[ht]
\begin{center}
    \if1\hidefigure
{
    \includegraphics[width=0.9\columnwidth]{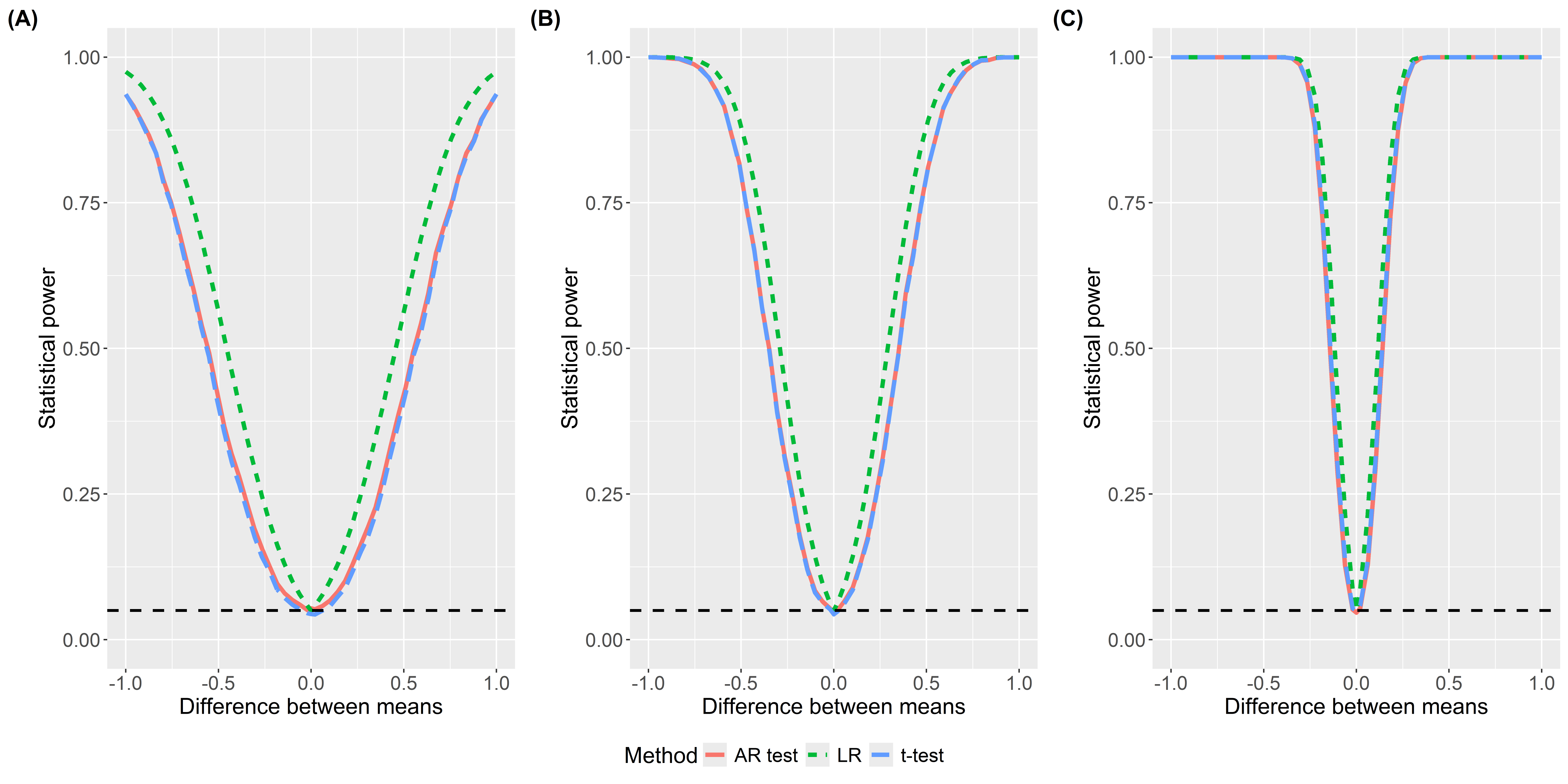}
}\fi
\caption{{Difference between two group means vs. the empirical power of the AR test (solid red line), Likelihood Ratio (LR) test (dashed green line), and Student's $t$-test (dashed blue line) as a function of the true difference between group means when the sample size changes: \textbf{(A)} $n = 26$; \textbf{(B)} $n = 64$; and \textbf{(C)} $n = 394$. The predetermined significance level 0.05 is illustrated with horizontal dashed line.
{\label{fig:effect_vs_power}}%
}}
\end{center}
\end{figure}

The power estimates of the AR test are comparable to those of the two-sample $t$-test when the groups are independent. Figure \ref{fig:effect_vs_power} shows how the empirical power of independent sample tests change when the difference between group means change. The AR test shows practically equal statistical power compared to the $t$-test although the $t$-test used here assumes equal variance corresponding to the ground truth population parameter values.

\subsection{Testing if a population mean vector equals specific fixed vector}

Let $\mu_0 = (\mu_0^1, \ldots, \mu_0^p)^\top$ denote the assumption about the population mean vector $\mu$. Consider testing

\begin{align}
    H_0&: \mu = \mu_0 \nonumber \\
    H_A&: \mu_j \neq \mu_j^0 \text{\, for some\,} j \in \{1,\ldots, p\}. \nonumber
\end{align}

Like in the previous example, the ratio of pdf of the multivariate normal distribution and the multivariate $t$-distribution is used to built the AR test statistic when $\textbf{t}$ is defined as in (\ref{eq:mvclt}) and $T(\textbf{t}) = \text{I}[f_n(\textbf{t}\mid \boldsymbol{0}, \widehat{\Sigma})/g_n(\textbf{t} \mid \boldsymbol{0}, \widehat{\Sigma})]$, when $g_n(\textbf{t} \mid \boldsymbol{0}, \Sigma) \neq 0$. In this case, the AR test statistic $\rho(\textbf{X})$ can be computed as defined in (\ref{eq:rhoteststat}). In this example, both the sample covariance matrix and the true population covariance matrix $\Sigma$ are used in the computation of the AR test statistic.

\begin{figure}[ht]
\begin{center}
    \if1\hidefigure
{
    \includegraphics[width=0.9\columnwidth]{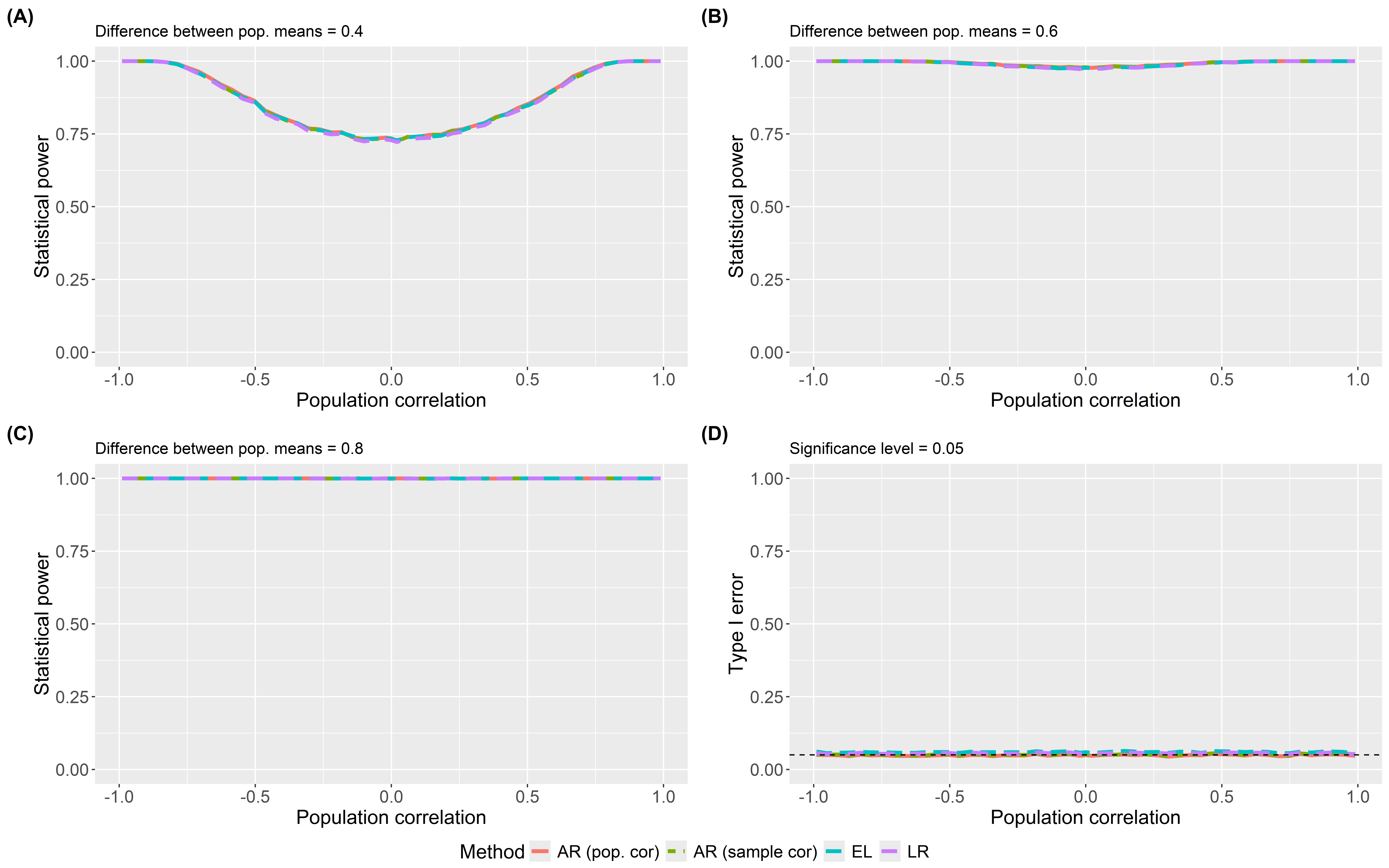}
}\fi
\caption{{Population correlation vs. power function 
of the AR test using the population covariance (AR pop. cor) (red solid line), AR test using the sample covariance (AR sample cor) (green dashed line), empirical Likelihood Ratio test (EL) (dashed blue line), and Likelihood Ratio test (LR) (purple dashed line). Different panels illustrate the power vs. correlation when the true difference between two group means is: \textbf{(A)} 0.4; \textbf{(B)} 0.6; and \textbf{(C)} 0.8. Panel \textbf{(D)} illustrates the estimated Type I error associated with different tests. The predetermined significance level 0.05 is illustrated with horizontal dashed line in panel \textbf{(D)}. Here $n = 52$.
{\label{fig:fixed_norm_power}}%
}}
\end{center}
\end{figure}

In this example, $p = 2$. Two powerful tests are selected for comparison: the LR test and the empirical LR test (EL) \citep[see, e.g.,][]{Owen2001}. Their rejection regions are determined using the asymptotic distribution $\chi^2_2$. The EL test is computed with the R package \texttt{mvhtests}. The statistical power of the AR test, LR test, and EL is estimated as the ratio of significant tests among 10000 Monte Carlo samples. The empirical power of these tests is illustrated in Figure \ref{fig:fixed_norm_power}.

Figure \ref{fig:fixed_norm_power} shows how the statistical power of the AR test, LR test, and the EL test are practically identical and all tests are very powerful. There is no difference whether the sample covariance or the population covariance matrix is used to compute the test statistic $\rho(\textbf{X})$.

\subsection{A test for goodness-of-fit}\label{sec:goodness-of-fit}

Consider the investigation of whether a sample originates from a specific probability distribution. In this case, the hypothesis of the problem can be formulated as follows:

\begin{align}
H_0&: f(X_i) = f_0(X_i) \text{ for all } X_i\in\mathbb{R}^p, \nonumber \\
H_A&: f(X_i) \neq f_0(X_i) \text{ for some } X_i\in\mathbb{R}^p, \nonumber
\end{align}
where $f_0$ is the theoretical population density distribution.

Here, the goodness-of-fit type AR test described in Definition \ref{def:ar_stat_gof_e} and its expected value shown in theorem (\ref{theorm:rho_eq_e}) is employed to test the hypothesis described above. Let $\widehat{f}$ denote a density estimate constructed, for example, using kernel density estimation. Here, the R package \texttt{Rfast2} using a high-degree memory and computationally optimized kernel functions is used to compute the density estimates $\widehat{f}$ \cite{Tsgris&Papadakis2018}.

The AR test is compared with the Kolmogorov–Smirnov (KS) test \citep[see, e.g.,][]{Fasano&Franceschini1987, Connor.etal2023}, Cram\'er–von Mises (CVM) test \citep[see, e.g.,][]{Chiu&Liu2009}, and Anderson-Darling test (AD) \citep{Anderson&Darling1952}. While testing normality, a novel test proposed by \cite{Szekely&Rizzo2005} (energy test (E)) is used. The AD and the energy tests are computed with R packages \texttt{goftest} and \texttt{energy}, respectively.

\begin{figure}[ht]
\begin{center}
    \if1\hidefigure
{
    \includegraphics[width=0.9\columnwidth]{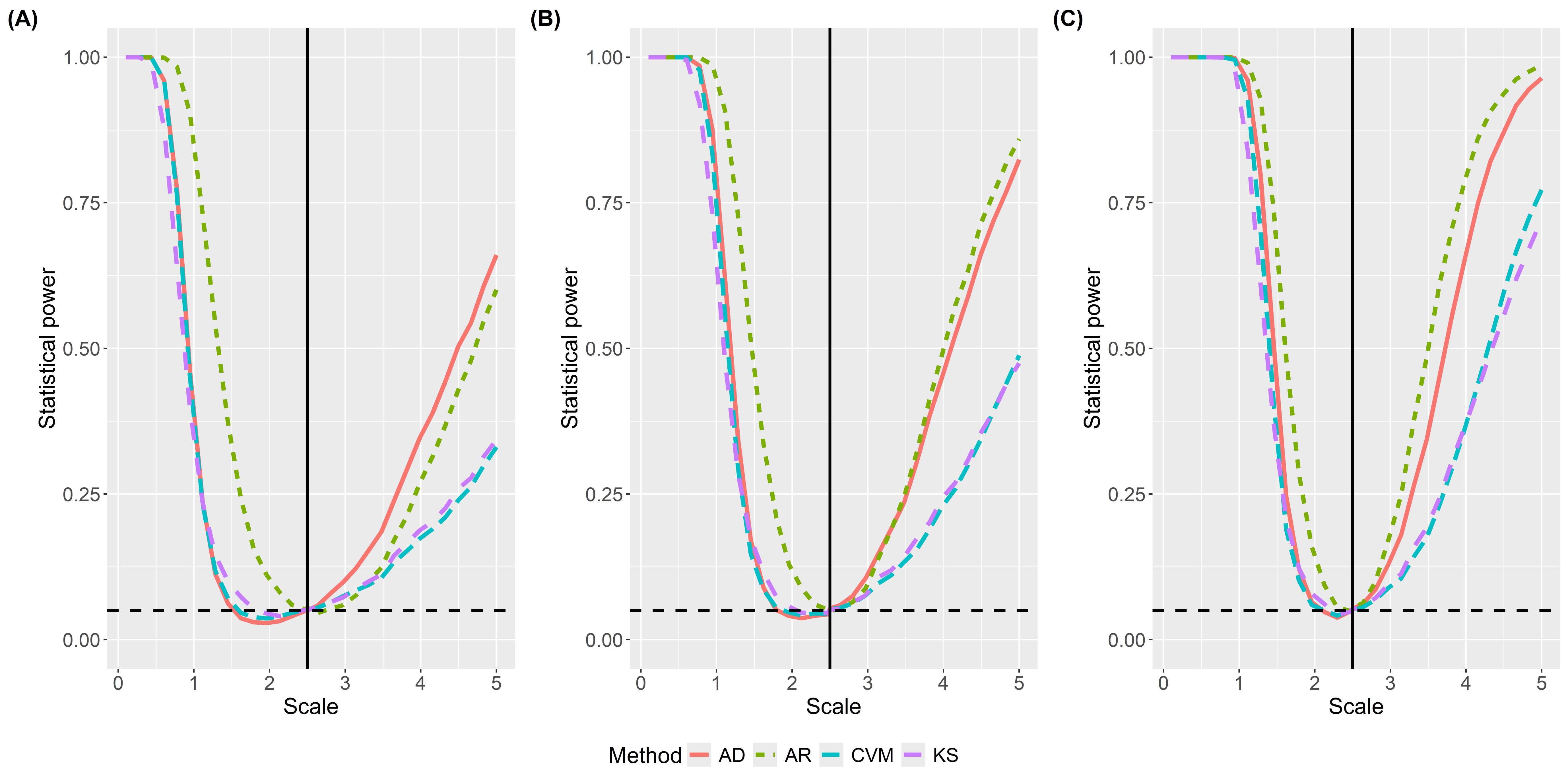}
}\fi
\caption{{Value of the scale parameter of the location-scale version of the t-distribution ($df=3$) vs. the empirical statistical power of the AR test (AR) (green dashed line), Cram\'er–von Mises test (CVM) (blue dashed line), the Kolmogorov–Smirnov test (KS) (purple dashed line), and Anderson-Darling test (AD) (solid red line). Different panels represent different sample sizes: \textbf{(A)} 20; \textbf{(B)} 30; and \textbf{(C)} 50. The vertical solid line corresponds to the value of the scale parameter under $H_0$, $\sigma_0 = 2.5$. The predetermined significance level $\alpha = 0.05$ is illustrated with horizontal dashed line.
{\label{fig:dist_power}}%
}}
\end{center}
\end{figure}

There are numerous distributions that can be tested using a goodness-of-fit test. Here it is investigated if a sample originate from a heavy-tailed univariate probability distribution with specific scale parameter value. In particular, it is examined wether the population distribution is a location-scale version of the $t$-distribution $t(df, \mu, \sigma)$ where $df$ is the number of degrees of freedom, $\mu$ is a location parameter, and $\sigma$ is a scale parameter. In this example, the null hypothesis is that the sample originates from $t(df=3, \mu=0, \sigma=\sigma_0)$ distribution, where $\sigma_0 = 2.5$. Three different sample sizes are considered: $n = 20, 30$, and  $50$. The empirical power of different tests is the ratio of significant tests of 10000 Monte Carlo samples at $\alpha = 0.05$. The power of the AR, CVM, KS, and AD tests are illustrated in Figure \ref{fig:dist_power}. 

The results summarized in Figure \ref{fig:dist_power} show that the power of the AR test is higher than that of the KS and CVM tests, even for the small sample sizes considered in this example. The power of the AR test is slightly higher than that of the AD test when $\sigma < \sigma_0$. As the sample size increases, the AR test appears to be the most powerful among those considered in this example.

The power of the AR test is also considered for assessing both univariate and multivariate normality. The null hypothesis is tested against four alternatives: the $t$-distribution $t(2)$, a normal mixture $0.5N(0, 1) + 0.5N(3, 1)$, the logistic distribution, and the standard uniform distribution. The empirical power of each test is defined as the ratio of significant tests over 10000 Monte Carlo samples at $\alpha = 0.05$. Estimates of the statistical power for the univariate normality test are summarized in Table \ref{tab:1} for sample sizes $20, 30$ and $50$.

\begin{table}[ht]
\begin{center}
\begin{tabular}{l|l|ccccc}
\hline
$n$ & Alternative distribution & AR & KS & CVM & E & AD \\ \hline
20 & $N(0, 1)$ & 0.04 & 0.05 & 0.05 & 0.05 & 0.05 \\ 
 & $t(2)$ & 0.27 & 0.09 & 0.10 & 0.51 & 0.60 \\
 & $0.5N(0,1)+0.5N(3,1)$ & 0.99 & 0.97 & 0.96 & 0.15 & 1.00 \\
 & logistic(0,1) & 0.65 & 0.23 & 0.23 & 0.10 & 0.72 \\
 & unif(0, 1) & 1.00 & 1.00 & 1.00 & 0.15 & 1.00 \\
 \hline 
30 & $N(0, 1)$ & 0.05 & 0.05 & 0.05 & 0.05 & 0.05 \\ 
 & $t(2)$ & 0.36 & 0.10 & 0.11 & 0.67 & 0.72 \\
 & $0.5N(0,1)+0.5N(3,1)$ & 1.00 & 1.00 & 1.00 & 0.24 & 1.00 \\
 & logistic(0,1) & 0.81 & 0.33 & 0.34 & 0.12 & 0.85 \\
 & unif(0,1) & 1.00& 1.00 & 1.00 & 0.28 & 1.00 \\ \hline
50 & $N(0, 1)$ & 0.05 & 0.05 & 0.05 & 0.05 & 0.05 \\ 
 & $t(2)$ & 0.55 & 0.14 & 0.14 & 0.85 & 0.86 \\
 & $0.5N(0,1)+0.5N(3,1)$ & 1.00 & 1.00 & 1.00 & 0.44 & 1.00 \\
 & logistic(0,1) & 0.96 & 0.52 & 0.57 & 0.14 & 0.96 \\ 
 & unif(0,1 ) & 1.00 & 1.00 & 1.00 & 0.53 & 1.00
\end{tabular}
\end{center}
\caption{{The empirical statistical power of the AR test (AR), the Kolmogorov–Smirnov test (KS), Cram\'er–von Mises test (CVM), energy test (E), and Anderson-Darling Test (AD) of univariate normality at $\alpha = 0.05$. Here $N(0,1)$ corresponds to the null distribution.} \label{tab:1}}
\end{table}

As shown in Table \ref{tab:1}, when the alternative distribution is the $t(2)$ distribution, the energy test and the AD test are more powerful than the AR test. Nevertheless, the AR test is still more powerful than KS and CVM tests. The AR test and the AD test achieve the highest statistical power when the alternative distribution is either a normal mixture or a logistic distribution. In the case of the logistic distribution, the empirical power of the AR test and the AD test is substantially higher than that of the KS, CVM, and energy tests.

When the AR test is applied as a test of multivariate normality, it is compared with five state-of-the-art multivariate procedures: the energy test, the test proposed by \cite{Zhou&Shao2014}, referred to here as the $T_n$ test, test proposed by Henze and Zirkler (HZ test) \citep{Henze&Zirkler1990}, Royston's H test \citep{Royston1982}, and test proposed by Villasenor-Alva and Gonz\'alez-Estrada (VA-GE test) \citep{Villasenor&Gonzalez2009}. The $T_n$ test, HZ test, Royston's H test, and VA-GE test are computed with the R package \texttt{mvnormalTest}. The multivariate counterparts of the univariate distributions considered in the previous examples are used with dimension $p = 3$. The empirical power of each test is defined as the ratio of significant tests over 1000 Monte Carlo samples at $\alpha = 0.05$. Estimates of the power for the multivariate normality test are reported in Table \ref{tab:2}.

\begin{table}[ht]
\begin{center}
\begin{tabular}{l|l|cccccc}
\hline
$n$ & Alternative distribution & AR & E & $T_n$ & HZ & H & VA-GE \\ \hline
20 & $N(\textbf{0}, I)$ & 0.04 & 0.05 & 0.06 & 0.04 & 0.08 & 0.06 \\
 & $t(2, \textbf{0}, I)$ & 0.11 & 0.84 & 0.82 & 0.78 & 0.83 & 0.64 \\
 & $0.5N(\textbf{0}, I) + 0.5N(\textbf{3}, I)$ & 0.99 & 0.13 & 0.09 & 0.18 & 0.19 & 0.03 \\
 & logistic(\textbf{0}, \textbf{1}) & 0.76 & 0.19 & 0.22 & 0.11 & 0.23 & 0.22 \\
 & unif(0,1)$^3$ & 1.00 & 0.06 & 0.27 & 0.12 & 0.37 & 0.30 \\ \hline
30 & $N(\textbf{0}, I)$ & 0.06 & 0.04 & 0.05 & 0.05 & 0.08 & 0.05 \\
 & $t(2, \textbf{0}, I)$ & 0.09 & 0.94 & 0.95 & 0.92 & 0.93 & 0.84 \\
 & $0.5N(\textbf{0}, I) + 0.5N(\textbf{3}, I)$ & 1.00 & 0.25 & 0.16 & 0.40 & 0.35 & 0.04 \\
 & logistic(\textbf{0}, \textbf{1}) & 0.90 & 0.24 & 0.30 & 0.16 & 0.26 & 0.29 \\
 & unif(0,1)$^3$ & 1.00 & 0.12 & 0.54 & 0.29 & 0.73 & 0.67 \\ \hline
50 & $N(\textbf{0}, I)$ & 0.07 & 0.05 & 0.05 & 0.05 & 0.06 & 0.05 \\
 & $t(2, \textbf{0}, I)$ & 0.09 & 1.00 & 1.00 & 0.99 & 0.99 & 0.98 \\
 & $0.5N(\textbf{0}, I) + 0.5N(\textbf{3}, I)$ & 1.00 & 0.54 & 0.35 & 0.81 & 0.70 & 0.04 \\
 & logistic(\textbf{0}, \textbf{1}) & 0.98 & 0.37 & 0.50 & 0.26 & 0.40 & 0.45 \\
 & unif(0,1)$^3$ & 1.00 & 0.38 & 0.89 & 0.62 & 0.99 & 0.98 \\
\hline
\end{tabular}
\end{center}
\caption{{The empirical statistical power of the AR test (AR), energy test (E), $T_n$ test, Henze and Zirkler test (HZ), Royston's H test (H), and Villasenor-Alva and Gonzalez-Estrada test (VA-GE) of multivariate normality ($p=3$) at $\alpha = 0.05$. Here $N(\textbf{0}, I)$ corresponds to the null distribution. Here $\textbf{0}$, $\textbf{1}$, $\textbf{3}$, and $I$ denotes $(0,0,0)^\top$, $(1,1,1)^\top$, $(3,3,3)^\top$, and the identity matrix of size $3$, respectively.} \label{tab:2}}
\end{table}

The results presented in Table \ref{tab:2} indicate that the power of the AR test is lower than that of the other tests when the alternative distribution is the multivariate $t$-distribution. In all other cases, the AR test is found to be the most powerful among the tests of multivariate normality considered.

All of these examples show that the AR test maintains the Type I error at a pre-specified level of significance. The power of the AR test depends on the density estimation method which is in line with theorem (\ref{theorm:rho_eq_tvd}). For example, instead of using the kernel density estimator from the R package \texttt{Rfast2}, one viable option is to employ the vine-copula-based kernel density estimator of \cite{Nagler&Czado2016}, which is considered an efficient method for kernel density estimation in high-dimensional settings. Using this estimator improves the power of the AR test when the alternative distribution is a heavy-tailed t-distribution. However, in the other cases considered here, the statistical power decreases slightly (results not shown).

\section{Applications}

In this section, the tests derived with the AR method are applied to two real-world data sets: (1) an Amyloid-beta (A$\beta$) data set linked to Alzheimer's disease \citep{Pivtoraiko.etal2015}; and (2) a reaction time data set from Experiment 1 of \citet{Wagenmakers.etal2008}. These data sets are publicly available in the R packages \texttt{Stat2Data} and \texttt{rtdists}, respectively.

\subsection{Amyloid-beta}

Amyloid-$\beta$ (A$\beta$) is a protein fragment central to the pathogenesis of Alzheimer's disease \citep{Brien&Wong2011}. The data consist of A$\beta$ measurements (pmol/g) from frozen posterior cingulate cortex gray matter harvested at autopsy from Catholic priests \citep{Pivtoraiko.etal2015}. Cognitive impairment status is also recorded, with subjects classified into three groups: (1) no cognitive impairment before death (NCI, $n = 19$); (2) mild cognitive impairment (MCI, $n = 21$); and (3) mild to moderate Alzheimer's disease (mAD, $n = 17$). The AR test (\ref{eq:rhoteststat}) is applied to assess whether A$\beta$ levels differ among the NCI, MCI, and mAD groups. Because the data do not contain repeated measurements, $\widehat{\Sigma} = \text{diag}(\widehat{\sigma}_{\text{NCI}}^2, \widehat{\sigma}_{\text{MCI}}^2, \widehat{\sigma}_{\text{mAD}}^2)$ is used as an estimator of the population covariance matrix, where $\widehat{\sigma}_{\text{NCI}}^2$, $\widehat{\sigma}_{\text{MCI}}^2$, and $\widehat{\sigma}_{\text{mAD}}^2$ are the empircal variances of A$\beta$ measurements in the NCI, MCI, and mAD groups, respectively.

The value of the AR test statistic is $0.417$, with a corresponding Monte Carlo $p$-value of approximately $0.005$. The $95\%$ credible interval for the statistic $T(\textbf{X})$, computed from the percentiles of the $\text{Bin}(21, 0.417)$ distribution, is $[0.190; 0.619]$. Taken together, these results indicate that A$\beta$ levels differ significantly in at least one group compared to the others. Pairwise comparisons summarized in Table \ref{tab:3} show that there is a statistically significant difference between the A$\beta$ levels of the NCI and mAD, and MCI and mAD groups at $\alpha = 0.05$.

\begin{table}[ht]
\begin{center}
\begin{tabular}{l| c c c c}
\hline
Group Comparison & Diff. in Means & $T(\textbf{X})$ & 95\% CI of $T(\textbf{X})$ & $p$-value \\
\hline
NCI - mAD & -425.03 & 0.589 & [0.381; 0.810] & 0.018 \\
MCI - mAD & -420.25 & 0.553 & [0.333; 0.762] & 0.009 \\
NCI - MCI & -4.78 & 1.00 & [1.00; 1.00] & 1.00 \\
\end{tabular}
\end{center}
\caption{{Pairwise differences in mean A$\beta$ levels between cognitive status groups (NCI, MCI, mAD), along with the corresponding AR test statistics, $95\%$ credible intervals (CIs) of $T(\textbf{X})$, and Bonferroni-adjusted Monte Carlo $p$-values.} \label{tab:3}}
\end{table}

\subsection{Reaction times}

Reaction time (RT), typically measured in milliseconds as the time required to complete a task, is a widely used dependent variable in psychology \citep[see, e.g.,][]{Whelan2008, Boeck&Jeon2019}. RT distributions are almost always right-skewed, and thus asymmetric, heavy-tailed distributions are commonly employed to capture their shape. In this example, the AR goodness-of-fit test is applied to RT data from Experiment 1 of \citet{Wagenmakers.etal2008}. In this experiment, participants were instructed to respond either as quickly or as accurately as possible. Here, the analysis focuses on the distribution of RTs from the ``quick'' responses.

\begin{figure}[ht]
\begin{center}
    \if1\hidefigure
{
    \includegraphics[width=0.9\columnwidth]{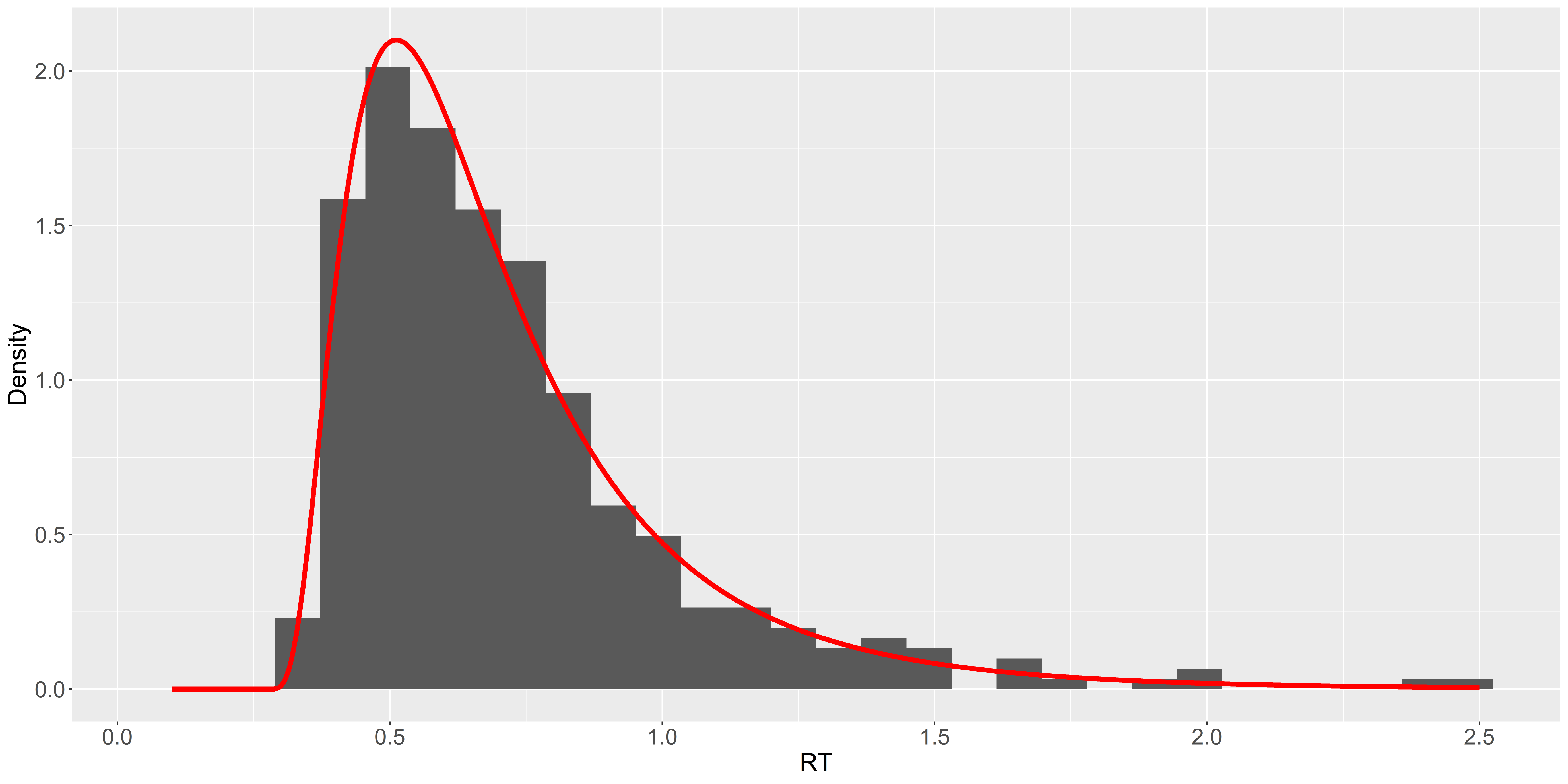}
}\fi
\caption{{Histogram of the RTs for participant ID 1, along with the fitted shifted log-normal distribution (solid red line), and normal distribution (dashed blue line).
{\label{fig:RT_dist}}%
}}
\end{center}
\end{figure}

Because a detailed analysis of the RTs does not contribute directly to this work, only the RTs from the participant with ID 1 are analyzed. Censored samples (uninterpretable responses, RT $<180$ ms, or RT $>3$ sec) and duplicated values were removed prior to analysis ($n=366$). A visual inspection of the data suggests that a shifted log-normal distribution provides a good fit to the reaction times. Figure \ref{fig:RT_dist} displays the histogram of the data along with the fitted shifted log-normal distribution, where the mean, standard deviation, and shift parameters were estimated using the R package \texttt{EnvStats}, \hl{as well as the fitted normal distribution, for which the mean and standard deviation were estimated from the data}.

The value of the AR test statistic is $0.961$ under $H_0$ that the RTs follow a shifted log-normal distribution with the estimated parameter values. \hl{The AR test statistic is $0.758$ under $H_0$ that the RTs follow a normal distribution}. The corresponding Monte Carlo $p$-value is approximately $0.894$ and $0.001$, respectively. The $95\%$ credible interval for the statistic $T(\textbf{X})$, computed from the percentiles of the Poisson binomial distribution (see (\ref{eq:poi_bin})) is $[0.940; 0.978]$ and $[0.713; 0.801]$, respectively. Taken together, these results suggest that the shifted log-normal distribution provides a close fit to the RTs considered in this example, \hl{offering a substantially better approximation than the normality assumption. Note that, although using estimated parameter values in hypothesis testing is questionable, this example illustrates how the AR test can be applied to evaluate different distributional assumptions rather than relying solely on the visual inspection of the data.}.

\section{Discussion}

The AR algorithm possesses untapped potential for identifying powerful statistical tests. The AR tests illustrated in this paper achieve statistical power comparable to that of UMP tests commonly used in the literature. In particular, the goodness-of-fit test derived from the AR framework appears to outperform some state-of-the-art methods. The AR test is general and applicable to arbitrary dimensions, providing an easy-to-use test with straightforward interpretation and connection to total variation distance.

This paper only scratched the surface of the potentials of the method. With minor modifications the AR test can be utilized in many statistical inference problems, such as mixed data, categorical variables, the two-sample problem, the $K$-sample problem, among others. The introduction of these extensions is left for future research.

There are numerous factors that can affect the statistical power and Type I error of any test, including unequal or extremely small sample sizes, missing data, unequal group variances, skewness and kurtosis of underlying distributions, and the curse of dimensionality, to name just a few. Researchers encounter these challenges frequently in real-world data analysis. These important questions will be addressed in future studies.

%\section*{Acknowledgements}\label{sec:acknowledgements}

\section*{Conflict of interest}

The author declares no conflict of interest.

\section*{Data availability}

The code to reproduce the examples, figures, and tables in this article are publicly available on GitHub at https://github.com/markkukuismin/Using-the-rejection-sampling-for-finding-tests-supplementary.

The Amyloid-$\beta$ data and the reaction time data are publicly available in the R packages \texttt{Stat2Data} and \texttt{rtdists}, respectively.

\begin{appendices}
\renewcommand{\thesection}{\Alph{section}.} % Add dot to the appendix name

\section{Proof of theorem \ref{theorm:rho_eq_e}}

\begin{proof}
    The proof of theorem \ref{theorm:rho_eq_e} closely follows \cite{Kuismin2025}. The expected value of $T(\textbf{X})$ follows from the properties of the indicator function, a uniformly distributed random variable, and the linearity of expectation. Let $I(x)$ be the indicator function of the event $x > U$ where $U \sim Unif(0, 1)$. Then $E[\text{I}(x)] = P(x > U) = 1$ if $x \geq 1$, and $P(x > U) = x$ if $0 < x < 1$. Let $b_i = f_0(X_i)/\widehat{f}(X_i)$. Let $S_1 = \{b_i \mid b_i \geq 1\}$ and $S_2 = \{b_i \mid b_i < 1\}$ denote sets that contain values of ratios $b_i$, $i = 1, \ldots, n$ which are greater than one or smaller than one, respectively. Let $\#S_1$ denote the cardinality of the set $S_1$. From this it follows that
    
\begin{align*}
    \rho(\textbf{X}) = E_U\left[T(\textbf{X})\right] &= E_U \Bigl\{n^{-1}\sum_{i=1}^n \text{I}[f_0(X_i)/\widehat{f}(X_i)]\Bigr\} = n^{-1}\sum_{i=1}^n P\Bigl[f_0(X_i)/\widehat{f}(X_i) > U\Bigr] \\
    &= n^{-1} \left[\#S_1 + \sum_{s_2 \in S_2} s_2 \right],
\end{align*}
This expectation can be rewritten as,

\begin{align*}
    \rho(\textbf{X}) = E_U\left[T(\textbf{X})\right] = n^{-1}\sum_{1=1}^n \min\left(1, f_0(X_i)/\widehat{f}(X_i)\right),
\end{align*}
which completes the proof.   
\end{proof}

\section{Proof of theorem \ref{theorm:rho_eq_tvd}}

\begin{proof}
In the following proof, it is assumed that the sample is drawn from a continuous population distribution with probability density function $f$, and that the distribution under the null hypothesis is also continuous with probability density function $f_0$. Both $f$ and $f_0$ are assumed to be strictly positive and bounded away from both zero and infinity on their support.

Let $X_1,\dots,X_n \in \mathbb{R}^p$ be i.i.d. random vectors with common density $f$ and $\textbf{X} = (X_1,\dots,X_n)$. Since $U_i \sim \text{Unif}(0,1)$ are independent of the data,

$$
E_U\left[\text{I}\left(\frac{f_0(X_i)}{\widehat{f}(X_i)} > U_i\right)\,\middle|\, \textbf{X}\right]= \min\!\left(1,\frac{f_0(X_i)}{\widehat{f}(X_i)}\right).
$$
Therefore,

$$
\rho(\textbf{X}) = E_U[T(\textbf{X}) \mid \textbf{X}] = \frac{1}{n} \sum_{i=1}^n \min\left(1,\frac{f_0(X_i)}{\widehat{f}(X_i)}\right).
$$
By the weak law of large numbers,

$$
\rho(\textbf{X})  \xrightarrow{P} E_f \left[\min\left(1,\frac{f_0(X)}{f(X)}\right)\right].
$$
Assume that the density estimator $\widehat{f}(X)$ converges to $f(X)$ in probability,

$$
\sup_{X \in\mathcal{X}} |\widehat{f}(X) - f(X)| \xrightarrow{P} 0.
$$
Since $f$ and $f_0$ are strictly positive and bounded away from zero and infinity on the support, the mapping

$$
g \mapsto \min\left(1,\frac{f_0(X)}{g(X)}\right)
$$
is continuous and uniformly Lipschitz in a neighborhood of $g=f$. Hence,

$$
E_f\left|\min\left(1,\frac{f_0(X)}{\widehat{f}(X)}\right)-\min\left(1,\frac{f_0(X)}{f(X)}\right)\right|\xrightarrow{P} 0,
$$
which implies

$$
E_f\left[\min\left(1,\frac{f_0(X)}{\widehat{f}(X)}\right)\right] 
\xrightarrow{P}
E_f\left[\min\left(1,\frac{f_0(X)}{f(X)}\right)\right].
$$
Because $\min(1, f_0/f) = \min(f_0,f)/f$, the expectation can be written as,

$$
E_f\left[\min\left(1,\frac{f_0(X)}{f(X)}\right)\right] = \int_{\mathcal{X}} \min(f_0(x),f(x))dx.
$$
By Scheffé’s Theorem \cite[see, e.g.,][]{Tsybakov2009}, integrating with respect to $X$ yields

$$
\int_{\mathcal{X}} \min(f_0(x), f(x)) dx = 1 - \frac{1}{2}\int_{\mathcal{X}} |f(x) - f_0(x)|dx,
$$
where $\frac{1}{2}\int_{\mathcal{X}} |f(x) - f_0(x)|dx$ is the total variation distance of probability density functions $f(x)$ and $f_0(x)$. Finally, one can write

$$
\rho(\textbf{X}) \xrightarrow{P} 1 - \|f(x) - f_0(x)\|_{\text{TV}},
$$
which completes the proof.
\end{proof}

\end{appendices}

\bibliographystyle{apalike}
\bibliography{references.bib}

@article{Anderson&Darling1952,
  title={Asymptotic theory of certain "goodness of fit" criteria based on stochastic processes},
  author={Anderson, Theodore W and Darling, Donald A},
  journal={The annals of mathematical statistics},
  pages={193--212},
  year={1952},
  volume = {23},
  note = {https://doi.org/10.1214/aoms/1177729437}
}

@book{Tsybakov2009,
  author       = {Alexandre B. Tsybakov},
  title        = {Introduction to Nonparametric Estimation},
  series       = {Springer Series in Statistics},
  publisher    = {Springer New York},
  address      = {New York, NY, USA},
  year         = {2009},
  edition      = {1st},
  isbn         = {978-0-387-79051-0, 978-0-387-79052-7},
  doi          = {10.1007/b13794},
  url          = {https://doi.org/10.1007/b13794}
}

@article{Eguchi&Copas2006,
title = {{Interpreting Kullback–Leibler divergence with the Neyman–Pearson lemma}},
journal = {Journal of Multivariate Analysis},
volume = {97},
pages = {2034--2040},
year = {2006},
note = {Special Issue dedicated to Prof. Fujikoshi},
doi = {https://doi.org/10.1016/j.jmva.2006.03.007},
author = {Shinto Eguchi and John Copas}
}

@article{Bodnar.etal2025,
author = {Bodnar, Taras and Parolya, Nestor and Veldman, Frederik},
title = {Nonlinear shrinkage test on a large-dimensional covariance matrix},
journal = {Statistica Neerlandica},
volume = {79},
pages = {e12348},
note = {https://doi.org/10.1111/stan.12348},
year = {2025}
}

@article{Brien&Wong2011,
   author = "Richard J. Brien and Philip C. Wong",
   title = "Amyloid Precursor Protein Processing and Alzheimer's Disease", 
   journal= "Annual Review of Neuroscience",
   year = "2011",
   volume = "34",
   pages = "185--204",
   note = "https://doi.org/10.1146/annurev-neuro-061010-113613"
  }

@article{Chiu&Liu2009,
title = {{Generalized Cram\'er–von Mises goodness-of-fit tests for multivariate distributions}},
journal = {Computational Statistics \& Data Analysis},
volume = {53},
pages = {3817--3834},
year = {2009},
note = {https://doi.org/10.1016/j.csda.2009.04.004},
author = {Sung Nok Chiu and Kwong I. Liu}
}

@article{Chen.etal2025,
author = {Chen, Yanzhou and Ding, Tianxuan and Wang, Xiufang and Zhang, Yaowu},
title = {A robust and powerful metric for distributional homogeneity},
journal = {Statistica Neerlandica},
volume = {79},
pages = {e12370},
note = {https://doi.org/10.1111/stan.12370},
year = {2025}
}

@book{Cohen1988,
  title={{Statistical Power Analysis for the Behavioral Sciences}},
  author={Jacob Cohen},
  year={1988},
  publisher={Routledge},
  edition = {2},
  note = {https://doi.org/10.4324/9780203771587}
}

@article{Fasano&Franceschini1987,
    author = {Fasano Giovanni and Ana Franceschini},
    title = {A multidimensional version of the Kolmogorov–Smirnov test},
    journal = {Monthly Notices of the Royal Astronomical Society},
    volume = {225},
    pages = {155--170},
    year = {1987},
    note = {https://doi.org/10.1093/mnras/225.1.155}
}

@article{Nagler&Czado2016,
title = {Evading the curse of dimensionality in nonparametric density estimation with simplified vine copulas},
journal = {Journal of Multivariate Analysis},
volume = {151},
pages = {69--89},
year = {2016},
note = {https://doi.org/10.1016/j.jmva.2016.07.003},
author = {Thomas Nagler and Claudia Czado}
}

@book{Robert&Casella2010,
 title={{Introducing Monte Carlo Methods with R}},
 author={Christian P. Robert and George Casella},
 year={2010},
 publisher={Springer, New York},
 note = {https://doi.org/10.1007/978-1-4419-1576-4},
 edition = {1st}
}

@book{Casella&Berger2024,
    author = {George Casella and Roger L. Berger},
    title = {{Statistical Inference}},
    publisher = {Chapman and Hall/CRC},
    year = 2024,
    note = {https://doi.org/10.1201/9781003456285},
    edition = {2nd}
}

@article{Henze&Zirkler1990,
author = {N. Henze and B. Zirkler},
title = {A class of invariant consistent tests for multivariate normality},
journal = {Communications in Statistics - Theory and Methods},
volume = {19},
pages = {3595--3617},
year = {1990},
note = {https://doi.org/10.1080/03610929008830400}
}

@article{Howes2023,
  title={{Markov chain Monte Carlo Significance tests}},
  author={Michael Howes},
  journal={arXiv preprint arXiv:2310.04924},
  year={2023}
}

@book{Kanji2006,
    author = {Gopal K. Kanji},
    title = {{100 Statistical Tests}},
    publisher = {London: SAGE Publications Ltd},
    year = {2006},
    edition = {3rd},
    note = {https://doi.org/10.4135/9781849208499}
}

@article{Kuismin2025,
author = {Markku Kuismin},
title = {Using Rejection Sampling Probability of Acceptance as a Measure of Independence},
journal = {Journal of Computational and Graphical Statistics},
volume = {34},
pages = {759--770},
year = {2025},
note = {https://doi.org/10.1080/10618600.2024.2388544}
}

@book{Lehmann&Romano2005,
  title={Testing statistical hypotheses},
  author={Erich L. Lehmann and Joseph P. Romano},
  edition={3rd},
  year={2005},
  note = {https://doi.org/10.1007/0-387-27605-X},
  publisher={Springer, New York}
}

@article{Lewandowski.etal2009,
title = {Generating random correlation matrices based on vines and extended onion method},
journal = {Journal of Multivariate Analysis},
volume = {100},
pages = {1989--2001},
year = {2009},
note = {https://doi.org/10.1016/j.jmva.2009.04.008},
author = {Daniel Lewandowski and Dorota Kurowicka and Harry Joe},
}

@article{Connor.etal2023,
  author = {Connor Puritz and Elan Ness-Cohn and Rosemary Braun},
  title = {{\texttt{fasano.franceschini.test}: An Implementation of a Multivariate KS Test in R}},
  journal = {The R Journal},
  year = {2023},
  note = {https://doi.org/10.32614/RJ-2023-067},
  volume = {15},
  pages = {159 -- 171}
}

@article{Pivtoraiko.etal2015,
title = {{Cortical pyroglutamate amyloid-$\beta$ levels and cognitive decline in Alzheimer's disease}},
journal = {Neurobiology of Aging},
volume = {36},
pages = {12--19},
year = {2015},
note = {https://doi.org/10.1016/j.neurobiolaging.2014.06.021},
author = {Violetta N. Pivtoraiko and Eric E. Abrahamson and Sue E. Leurgans and Steven T. DeKosky and Elliott J. Mufson and Milos D. Ikonomovic}
}

@article{Reshef.etal2018,
  author = {David N. Reshef and Yakir A. Reshef and Pardis C. Sabeti and Michael Mitzenmacher},
  title = {{An empirical study of the maximal and total information coefficients and leading measures of dependence}},
  volume = {12},
  journal = {The Annals of Applied Statistics},
  pages = {123 -- 155},
  year = {2018},
  note = {https://doi.org/10.1214/17-AOAS1093}
}

@article{Royston1982,
    author = {Royston, J. P.},
    title = {{An extension of Shapiro and Wilk’s W test for normality to large samples}},
    journal = {Journal of the Royal Statistical Society Series C: Applied Statistics},
    volume = {31},
    pages = {115--124},
    year = {1982},
    note = {https://doi.org/10.2307/2347973}
}

@article{Student1908,
 author = {Student},
 journal = {Biometrika},
 pages = {1--25},
 title = {The Probable Error of a Mean},
 note = {https://doi.org/10.2307/2331554},
 volume = {6},
 year = {1908}
}

@article{Szekely&Rizzo2005,
title = {A new test for multivariate normality},
journal = {Journal of Multivariate Analysis},
volume = {93},
pages = {58--80},
year = {2005},
note = {https://doi.org/10.1016/j.jmva.2003.12.002},
author = {Gábor J. Székely and Maria L. Rizzo}
}

@article{Villasenor&Gonzalez2009,
author = {José A. Villasenor Alva and Elizabeth González Estrada},
title = {{A generalization of Shapiro–Wilk's test for multivariate normality}},
journal = {Communications in Statistics - Theory and Methods},
volume = {38},
pages = {1870--1883},
year = {2009},
note = {https://doi.org/10.1080/03610920802474465}
}

@book{Owen2001,
    author = {Art B. Owen},
    title = {Empirical Likelihood},
    edition = {1st},
    publisher = {Chapman and Hall/CRC.},
    year = 2001,
    note = {https://doi.org/10.1201/9781420036152}
}

@article{Tsgris&Papadakis2018,
    author = {Michail Tsagris and Manos Papadakis},
    title = {{Taking R to its limits: 70+ tips}},
    journal = {PeerJ Preprints},
    year = {2018},
    note = {https://doi.org/10.7287/peerj.preprints.26605v1},
    volume = {6},
    pages = {e26605v1}
    
}

@article{Wagenmakers.etal2008,
title = {A diffusion model account of criterion shifts in the lexical decision task},
journal = {Journal of Memory and Language},
volume = {58},
pages = {140--159},
year = {2008},
note = {https://doi.org/10.1016/j.jml.2007.04.006},
author = {Eric-Jan Wagenmakers and Roger Ratcliff and Pablo Gomez and Gail McKoon}
}

@article{Whelan2008,
  title={Effective analysis of reaction time data},
  author={Robert Whelan},
  journal={The Psychological Record},
  volume={58},
  pages={475--482},
  year={2008},
  note= {https://doi.org/10.1007/BF03395630}
}

@ARTICLE{Boeck&Jeon2019,
AUTHOR={Paul De Boeck and Minjeong Jeon},
TITLE={An Overview of Models for Response Times and Processes in Cognitive Tests},
JOURNAL={Frontiers in Psychology},
VOLUME={10},
YEAR={2019},
note = {https://doi.org/10.3389/fpsyg.2019.00102}
}

@article{Zhou&Shao2014,
    author = {Ming Zhou and Yongzhao Shao},
    title = {A Powerful Test for Multivariate Normality},
    journal = {Journal of applied statistics},
    year = {2014},
    volume = {41},
    pages = {351--363},
    note = {https://doi.org/10.1080/02664763.2013.839637}
}

\end{document}